\documentclass[prl,twocolumn,reprint,preprintnumbers,amsmath,amssymb]{revtex4-1}
\usepackage{graphicx}
\usepackage{dcolumn}
\usepackage{bm}
\usepackage{amssymb}
\usepackage{epsfig}
\begin{document}
\title{The Josephson heat interferometer}
\author{F. Giazotto}
\email{giazotto@sns.it}
\affiliation{NEST, Istituto Nanoscienze-CNR  and Scuola Normale Superiore, I-56127 Pisa, Italy}
\author{M. J. Mart\'inez-P\'erez}
\affiliation{NEST, Istituto Nanoscienze-CNR  and Scuola Normale Superiore, I-56127 Pisa, Italy}
\maketitle

\textbf{The Josephson effect \cite{Josephson} represents perhaps the prototype of macroscopic phase coherence and is at the basis of the most widespread interferometer, i.e., the superconducting quantum interference device (SQUID) \cite{Clarke}. 
Yet, in analogy to electric interference, 
Maki and Griffin \cite{Maki} predicted in 1965 that thermal current flowing through a temperature-biased Josephson tunnel junction is a stationary periodic function of the quantum phase difference between the superconductors.
The interplay between quasiparticles and Cooper pairs condensate is at the origin of such phase-dependent heat current, and is unique to Josephson junctions.
In this scenario, a temperature-biased SQUID would allow heat currents to interfere \cite{Guttman2,Giazotto} thus implementing the thermal version of the electric Josephson interferometer.
The dissipative character of heat flux makes this coherent phenomenon not less extraordinary than its electric (non-dissipative) counterpart.
Albeit weird, this striking effect has never been demonstrated so far. 
Here we report the first experimental realization of a heat interferometer.
We investigate heat exchange between two normal metal electrodes kept at different temperatures and tunnel-coupled to each other through a thermal `modulator' \cite{Giazotto} in the form of a DC-SQUID. 
Heat transport in the system is found to be phase dependent,
in agreement with the original prediction.
With our design the Josephson heat interferometer yields magnetic-flux-dependent temperature oscillations of amplitude up to $\sim 21$ mK, and provides a flux-to-temperature transfer coefficient exceeding $\sim60\text{mK}/\Phi_0$ at 235 mK ($\Phi_0\simeq 2\times 10^{-15}$ Wb is the flux quantum).    
Besides offering remarkable insight into thermal transport in Josephson junctions, our results represent a significant step toward phase-coherent mastering of heat in solid-state nanocircuits, and pave the way to the design of novel-concept coherent caloritronic devices.}
\begin{figure}[t!]
\includegraphics[width=\columnwidth]{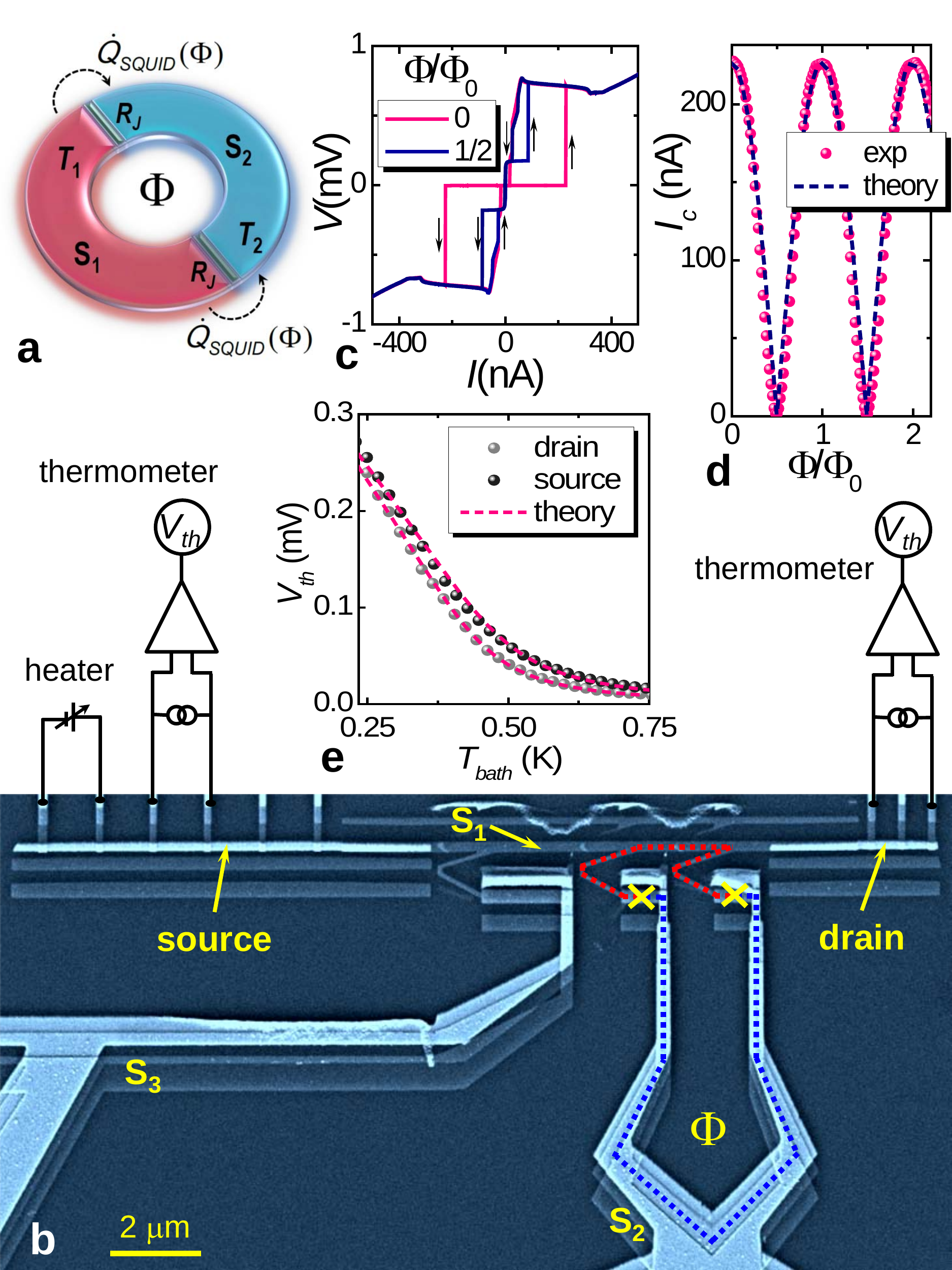}
\caption{\label{fig1} \textbf{Josephson heat interferometer.} 
\textbf{a}, Scheme of the device core:
two identical superconductors S$_1$ and S$_2$ kept at different temperatures $T_1$ and $T_2$ (with $T_1\geq T_2$), respectively, are tunnel coupled so to implement a DC-SQUID. 
$R_J$ is the normal-state resistance of each junction,
$\Phi$ is the applied magnetic flux threading the loop, whereas $\dot{Q}_{SQUID}(\Phi)$ is the heat current flowing from hotter to colder superconductor.
\textbf{b}, Scanning electron micrograph of the heat interferometer. Cu source and drain electrodes share the contact through AlO$_x$ tunnel barriers with an Al island (S$_1$)  defining one branch of a DC-SQUID. The other branch of the SQUID (S$_2$) extends into a large volume lead to insure proper thermalization of its quasiparticles at the bath temperature $T_{bath}$. SQUID junctions are marked by crosses. 
S$_1$ is also in contact with an Al tunnel probe (S$_3$) enabling independent characterization of the SQUID.
NIS junctions made of Al  are used as heaters and thermometers. 
See text for details. 
\textbf{c}, SQUID voltage ($V$) versus current ($I$) characteristics at two representative values of the applied flux. 
$\Phi_0$ corresponds to an applied field $B=\Phi_0/A\approx 1$ Oe, where $A\approx 19.6\,\mu$m$^2$ is the ring area. 
\textbf{d} $\Phi$-dependent pattern of the SQUID critical current $I_c$. Dashed line is the theoretical result for a DC-SQUID assuming $\sim 0.3\%$ asymmetry between the critical currents of the two junctions.
Data in \textbf{c} and \textbf{d} were taken at 240 mK.
\textbf{d}, Source and drain NIS thermometers calibration curves. Symbols indicate the measured voltage drop $V_{th}$ at a bias current of 70 pA (source) and 40 pA (drain)  versus $T_{bath}$. Dashed lines are theoretical results for a NIS junction. 
}
\end{figure}

To figure out a Josephson heat interferometer we consider a symmetric DC-SQUID (i.e., a superconducting loop comprising two equal Josephson tunnel junctions with resistance $R_J$) composed of two identical superconductors S$_1$ and S$_2$ in thermal equilibrium and residing at temperatures $T_1$ and $T_2$, respectively (see Fig. 1a). For definiteness, we assume $T_1\geq T_2$ so that the SQUID is temperature-biased only.
Within this assumptions, the total heat flow $\dot{Q}_{SQUID}$ from S$_1$ to S$_2$ becomes stationary and is given by \cite{Maki,Giazotto,Sauls1,Sauls2}
\begin{equation}
\dot{Q}_{SQUID}(\Phi)=2\dot{Q}_{qp}-2\dot{Q}_{int}\left|\text{cos}\left(\frac{\pi\Phi}{\Phi_0}\right)\right|,
\label{powerflow}
\end{equation} 
where the factor $2$ accounts for two identical SQUID junctions, 
and $\Phi$ is the applied magnetic flux threading the loop. 
$\Phi$-dependence appears in Eq. (\ref{powerflow}) only through the cosine term so that $\dot{Q}_{SQUID}$ consists of a $\Phi_0$-periodic function 
superimposed on top of a magnetic flux-independent component.
In the above expression,
$\dot{Q}_{qp}(T_1,T_2)=\frac{2}{e^2 R_{J}}\int^{\infty}_{0} d\varepsilon \varepsilon \mathcal{N}_1 (\varepsilon,T_1)\mathcal{N}_2 (\varepsilon,T_2)[f_1(\varepsilon,T_1)-f_2(\varepsilon,T_2)]$ is the usual  quasiparticles heat current \cite{Maki,RMP}, whereas $\dot{Q}_{int}(T_1,T_2)=\frac{2}{e^2 R_{J}}\int^{\infty}_{0} d\varepsilon \varepsilon \mathcal{M}_1 (\varepsilon,T_1)\mathcal{M}_2 (\varepsilon,T_2)[f_1(\varepsilon,T_1)-f_2(\varepsilon,T_2)]$ is the power flow due to interference between quasiparticles and the Cooper pairs condensate \cite{Maki,Guttman2,Guttman1,Sauls1,Sauls2}. 
$f_i(\varepsilon,T_i)=[1+\text{exp}(\varepsilon/k_BT_i)]^{-1}$ is the Fermi-Dirac distribution at temperature $T_i$ ($i=1,2$), $\mathcal{N}_{i}(\varepsilon,T_i)=|\varepsilon|/\sqrt{\varepsilon^2-\Delta_{i}(T_i)^2}\Theta[\varepsilon^2-\Delta_{i}(T_i)^2]$ is the BCS normalized density of states in the superconductors \cite{Tinkham}, 
$\mathcal{M}_{i}(\varepsilon,T_i)=\Delta_{i}(T_i)/\sqrt{\varepsilon^2-\Delta_{i}(T_i)^2}\Theta[\varepsilon^2-\Delta_{i}(T_i)^2]$ \cite{Maki,Guttman2,Guttman1,Sauls1,Sauls2}, 
$\Delta_i(T_i)$ is the temperature-dependent energy gap \cite{Tinkham},
$\Theta(x)$ is the Heaviside function,
$k_B$ is the Boltzmann constant, and $e$ is the electron charge.
We note that both $\dot{Q}_{qp}$ and $\dot{Q}_{int}$ vanish for $T_1= T_2 $, whereas $\dot{Q}_{int}$ also vanishes when at least one of the superconductors is in the normal state.

The implementation of our heat interferometer is shown in Fig. 1b. The structure has been fabricated with electron-beam lithography and three-angle shadow-mask evaporation. It consists of a source and drain copper (Cu) electrodes tunnel-coupled to a superconducting aluminum (Al) island (S$_1$)  defining one branch of a DC-SQUID. Source and drain junctions normal-state resistances are $R_{s}\simeq 1.5\,\text{k}\Omega$ and $R_{d}\simeq 1\,\text{k}\Omega$, respectively, whereas the resistance of each SQUID junction is $R_J\simeq 1.3\,\text{k}\Omega$. 
S$_1$ is also contacted by an extra Al probe (S$_3$) via a tunnel junction with normal-state resistance $R_p\sim 0.55\, \text{k}\Omega$, enabling independent characterization of the SQUID.
Both source and drain are tunnel-coupled to a few external Al probes (vertical wires in Fig. 1b) so to realize normal metal-insulator-superconductor (NIS) junctions, with normal-state resistance $\sim 25\,\text{k}\Omega$ each, which allow Joule heating and thermometry \cite{RMP}.

Below the critical temperature of Al ($\simeq 1.4$ K) Josephson coupling allows dissipationless charge transport through the SQUID. The SQUID voltage-current characteristics at $240$ mK for two representative magnetic-flux values are shown in Fig. 1c. 
In particular, a well-defined Josephson current with maximum amplitude of $\simeq226$ nA is observed at $\Phi=0$.
The magnetic-flux pattern of the SQUID critical current $I_c$ along with the theoretical prediction \cite{Clarke,Tinkham} is displayed in Fig. 1d, and shows a nearly-complete supercurrent modulation which confirms the good symmetry of the SQUID.

Thermal transport and, therefore, heat interference in the structure, arises from heating electrons in the source above lattice temperature ($T_{bath}$) so to elevate the quasiparticles temperature in S$_1$ ($T_1$) and
create a temperature gradient across the SQUID. 
This hypothesis is expected to hold as the second branch of the SQUID (S$_2$) extends into  a large volume lead providing efficient thermalization of its quasiparticles at $T_{bath}$.
$\dot{Q}_{SQUID}$ will thus manifest itself leading to a $\Phi_0$-periodic modulation of drain electron temperature ($T_{drain})$.

Investigation of heat transport in our system is performed as follows. One pair of NIS junctions in the source is operated as a heater whereas a second pair is used to measure electron temperature ($T_{source}$) by applying a small DC bias current and recording the  corresponding temperature-dependent voltage drop $V_{th}$ \cite{RMP,Nahum}. 
Analogously, another pair of NIS junctions is used to perform thermometry in the drain (see Fig. 1b).
Thermometers bias currents were optimized to achieve high sensitivity while limiting the impact of self-heating or self-cooling \cite{RMP}. 
Figure 1e displays the calibration curves of source and drain  thermometers versus $T_{bath}$, obtained by slowly sweeping the cryostat temperature from 235 mK to 750 mK. 
The corresponding theoretical results for a NIS junction \cite{Tinkham} are shown as well.

Measurement of heat interference is done by stabilizing the cryostat temperature at a desired $T_{bath}$ and heating the source up to a given $T_{source}$. $T_{drain}$ is then recorded  against a slowly-sweeping external magnetic flux.
Figure 2a shows $T_{drain}$ against $\Phi$ measured at 235 mK for increasing values of $T_{source}$. 
Notably, $T_{drain}$ is $\Phi_0$-periodic  in $\Phi$, as the Josephson critical current (see Fig. 1d). 
As we shall argue, such a temperature modulation is of coherent nature, and stems from magnetic flux-control of $\dot{Q}_{SQUID}$ which is a hallmark of the Josephson effect. 
By raising $T_{source}$ leads to a monotonic enhancement of the average drain temperature over one flux quantum, $\left\langle T_{drain}\right\rangle$, which follows from increased heat flow across the structure. 
On the other hand, the modulation amplitude  $\delta T_{drain}$, defined as the difference between the maximum and minimum values of $T_{drain}$, turns out to initially increase and then tends to saturate
at larger $T_{source}$.
In particular, $\delta T_{drain}$ up to $\sim 21$ mK is observed corresponding to $\sim 9\%$ of relative modulation amplitude at 235 mK.
$T_{drain}$ modulation is observed also without intentional source heating,
and might be related to a parasitic power ($\sim1-5$ fW) in the structure.    
The full $T_{source}$-dependence of $\left\langle T_{drain}\right\rangle$ and $\delta T_{drain}$ are displayed in Fig. 2b and confirm the above described behavior. 
We stress that heat interference manifests itself only owing to the existence of a finite temperature bias across the SQUID.
Any voltage drop occurring at the Josephson junctions makes the phase-coherent component of  $\dot{Q}_{SQUID}$ time-dependent thus  not contributing to time-averaged heat transport \cite{Maki,Guttman2,Guttman1}.

A relevant figure of merit of the heat interferometer is represented by the flux-to-temperature transfer coefficient \cite{Giazotto}, $\mathcal{T}\equiv \partial T_{drain}/\partial \Phi$, shown in Fig. 2c versus $\Phi$ for a few selected $T_{source}$. 
It turns out that $|\mathcal{T}|$ exceeding $60\,\text{mK}/\Phi_0$ is obtained at $675\,\text{mK}$.  Larger values might be obtained by lowering $T_{bath}$ and by further optimizing the structure design.
\begin{figure}[t!]
\includegraphics[width=\columnwidth]{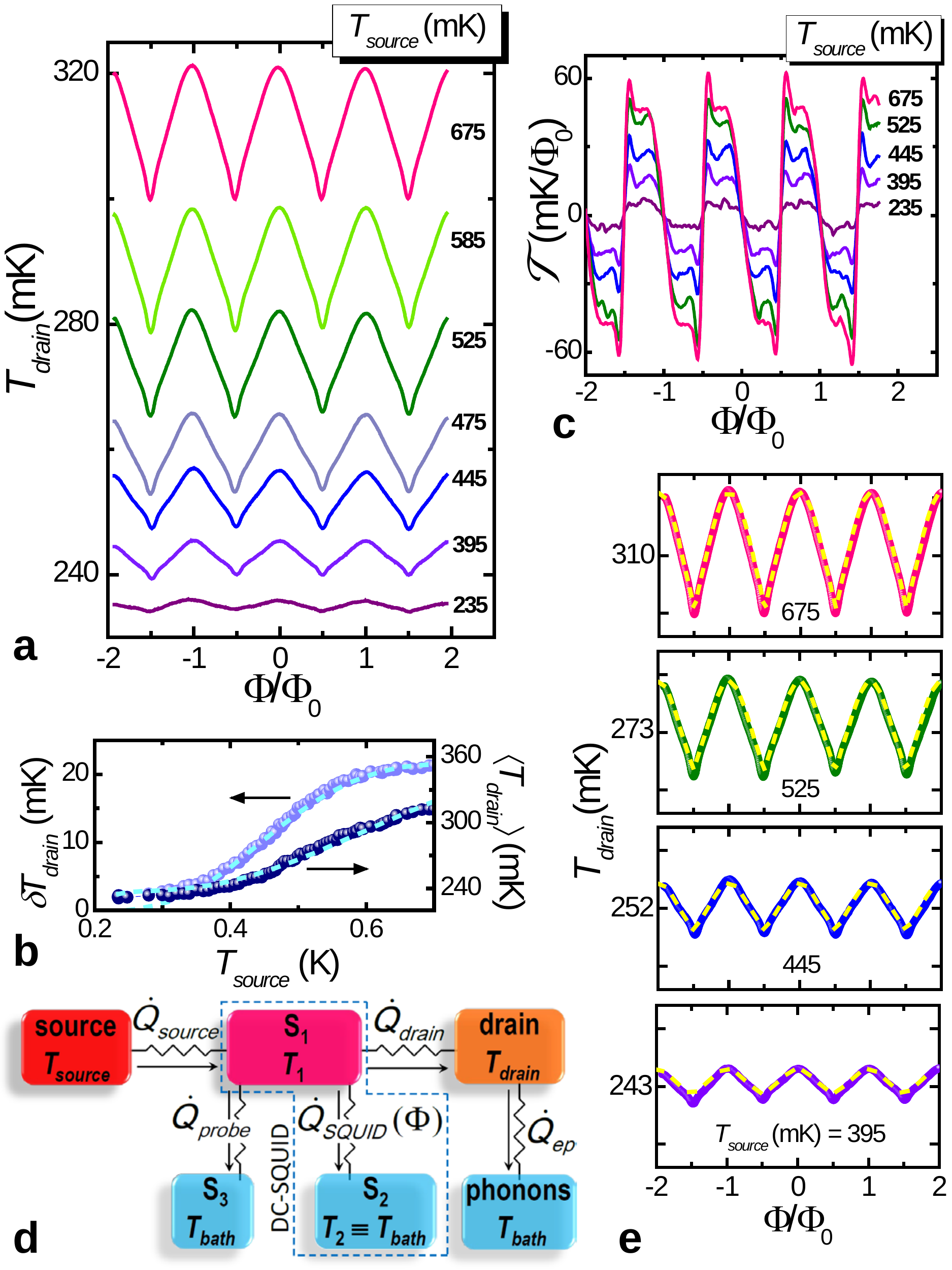}
\caption{\label{fig2} \textbf{Behavior of the heat interferometer at $235$ mK.}
\textbf{a}, Flux modulation of $T_{drain}$ measured for several $T_{source}$ values.
\textbf{b}, Modulation amplitude $\delta T_{drain}$ (left axis) and average temperature $\left\langle T_{drain}\right\rangle$ (right axis) versus $T_{source}$. Dashed lines are the results from thermal model (see discussion below and in the text).
\textbf{c}, Flux-to-temperature transfer function $\mathcal{T}\equiv \partial T_{drain}/\partial \Phi$ versus $\Phi$ measured at a few selected $T_{source}$.
\textbf{d}, Idealized thermal diagram accounting for our setup. S$_1$ exchanges energy at power $\dot{Q}_{source}$ and $\dot{Q}_{drain}$ due to quasiparticle heat conduction with source and drain, respectively, at power $\dot{Q}_{SQUID}$ with S$_2$ and $\dot{Q}_{probe}$ with S$_3$. S$_2$ and S$_3$
are assumed to reside at $T_{bath}$.
Drain electrons exchange energy at power $\dot{Q}_{drain}$ with S$_1$, and at power $\dot{Q}_{ep}$ with lattice phonons thermalized at bath temperature. Arrows indicate the direction of heat flows for $T_{bath}<T_{drain}<T_1<T_{source}$.
\textbf{e}, Experimental $T_{drain}(\Phi)$  curves at a few selected values of $T_{source}$ along with the results from thermal model (dashed lines). The full temperature range in each panel is 28 mK, and the vertical division corresponds to 10 mK.
}
\end{figure}

To account for our observations we have elaborated a thermal model sketched
in Fig. 2d. We assume S$_1$ to exchange heat at power $\dot{Q}_{source}$ and $\dot{Q}_{drain}$ due to quasiparticle heat conduction with source and drain, respectively, at power $\dot{Q}_{SQUID}$ with S$_2$ and $\dot{Q}_{probe}$ with S$_3$.
Both S$_2$  and S$_3$ are assumed to be thermalized at $T_{bath}$. 
Furthermore, drain electrons exchange energy at power $\dot{Q}_{drain}$ with S$_1$, and at power  $\dot{Q}_{ep}$ with lattice phonons residing at $T_{bath}$ \cite{RMP,Roukes}. 
The thermal steady-state of the system may be described by the energy-balance equations
\begin{eqnarray}
-\dot{Q}_{source}+\dot{Q}_{probe}+\dot{Q}_{SQUID}(\Phi)+\dot{Q}_{drain}=0 \nonumber\\
-\dot{Q}_{drain}+\dot{Q}_{ep}=0,
\label{balance}
\end{eqnarray}
where first equation accounts for thermal budget in S$_1$, while the second one describes heat exchange in the drain. 
$T_1$ and $T_{drain}$ can be determined under given conditions by numerically solving Eqs. (\ref{balance}) (see Methods Summary for further details).
The model neglects heat exchange with photons due to mismatched impedance \cite{Cleland,Pekola,Timofeev,Peltonen,Pascal}, electron-phonon coupling in S$_1$ owing to its reduced volume and low experimental $T_{bath}$ \cite{Timofeev2}, as well as phonon heat current \cite{Maki}.

Figure 2e shows the comparison between experiment and model displaying a few $T_{drain}(\Phi)$ traces (solid lines) along with the theoretical behavior (dashed lines). Analogously, dashed lines in Fig. 2b show the predicted $\left\langle T_{drain}\right\rangle$ and $\delta T_{drain}$ against $T_{source}$.
Although idealized the model provides reasonable agreement with our observations, and grasps the relevant physical picture at the origin of heat interference in our system. 

The dependence on bath temperature is shown in Fig. 3a (left panel) which displays $T_{drain}(\Phi)$  at a few increasing $T_{bath}$ for $T_{source}$ set around 700 mK. 
Right panel of Fig. 3a shows the curves obtained from the model at the corresponding $T_{bath}$. 
Besides leading to a monotonic enhancement of $\left\langle T_{drain} \right\rangle$, by increasing $T_{bath}$ yields suppression of $\delta T_{drain}$ and smearing of $T_{drain}(\Phi)$ \cite{Giazotto} which can be mainly ascribed to the enhancement of electron-phonon coupling in the drain \cite{RMP} as well as to the influence of 
thermal broadening. 
$\delta T_{drain}\sim 2.5$ mK is still observable at 450 mK, whereas the modulation disappears for $T_{bath}\gtrsim 500$ mK. 
We emphasize that the latter is substantially smaller than the temperature setting the disappearance of the Josephson effect in the SQUID ($\simeq 1.4$ K). 
Figures 3b-d  show  $\left\langle T_{drain} \right\rangle$, $\delta T_{drain}$, and the maximum of $|\mathcal{T}|$ versus $T_{source}$, respectively, recorded at the same $T_{bath}$ as in panel 3a, along with the lines obtained from thermal model.
Each of these quantities displays similar qualitative behavior at different bath temperatures, and a smoothing of their characteristics is observed by increasing $T_{bath}$. 
This is consistent with the picture provided by the model which captures the main features of the experimental data.

\begin{figure}[t!]
\includegraphics[width=\columnwidth]{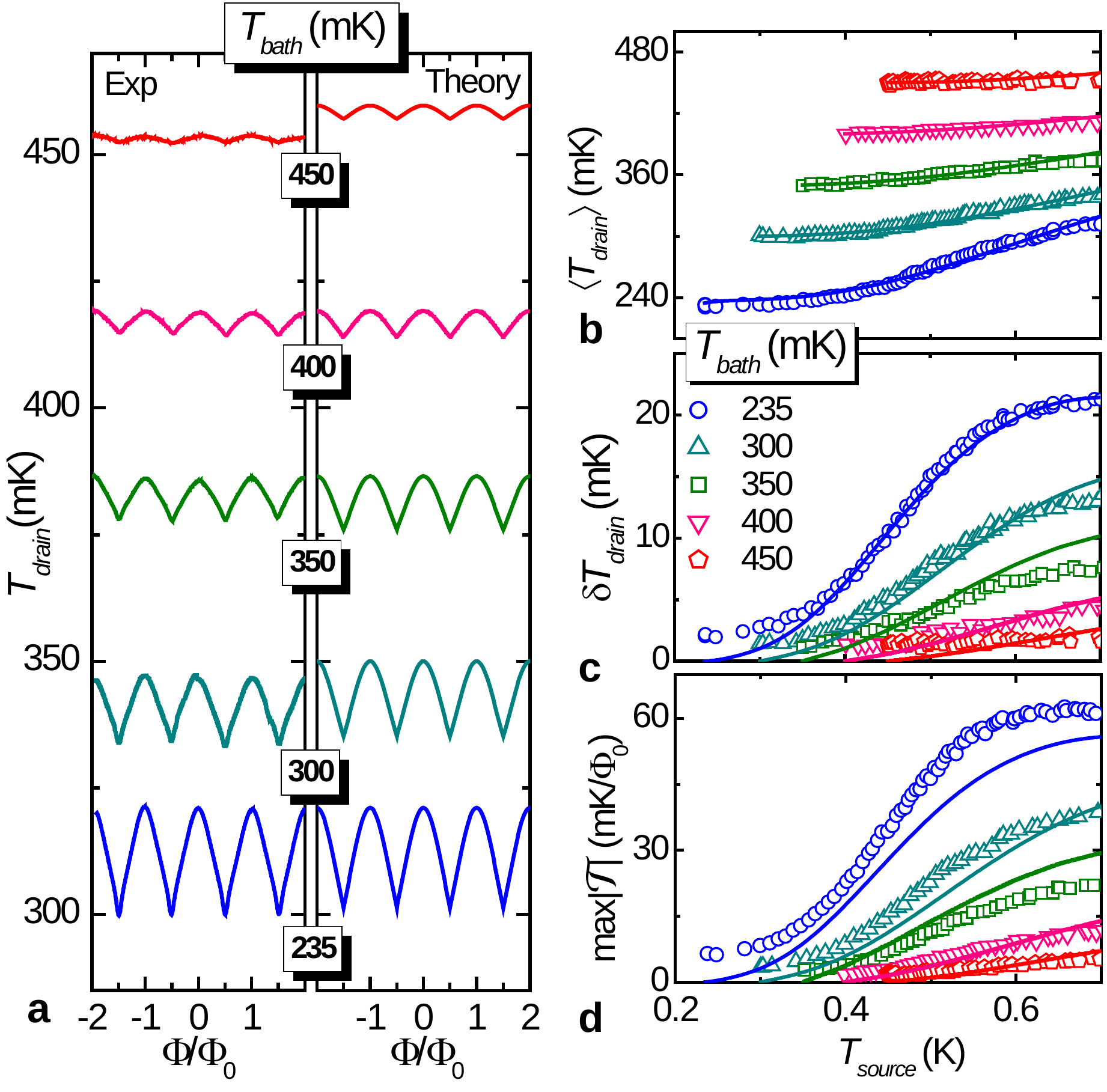}
\caption{\label{fig3} \textbf{Behavior of the heat interferometer at different bath temperatures.}
\textbf{a}, Flux modulation of $T_{drain}$ recorded at different $T_{bath}$. 
From bottom to top, data correspond to $T_{source}=675\,\text{mK},\,700\,\text{mK},\,690\,\text{mK},\,700\,\text{mK},\,700\,\text{mK}$.
Left panel shows the experimental data, while the right one displays results from the thermal model.  
A sizable temperature modulation is still observable at 450 mK, whereas $\delta T_{drain}$ vanishes for $T_{bath}\gtrsim 500$ mK.
\textbf{b}, Average temperature $\left\langle T_{drain}\right\rangle$ versus $T_{source}$. 
\textbf{c}, Modulation amplitude $\delta T_{drain}$ versus $T_{source}$.
\textbf{d}, Maximum value of $|\mathcal{T}|$ versus $T_{source}$.
Data in panels \textbf{b}-\textbf{d} were measured at the same $T_{bath}$ as in panel \textbf{a}. Solid lines correspond to the thermal model.
}
\end{figure}
Our results confirm what originally predicted almost fifty years ago \cite{Maki}, i.e., the existence of a phase-dependent component in the heat current flowing  through a temperature-biased Josephson tunnel junction. 
Besides offering insight into thermal transport in Josephson weak-links, this effect could represent a valuable tool toward phase-coherent manipulation of heat in solid-state nanocircuits \cite{RMP,DiVentra,Ojanen,Ruokola}. 
Yet, novel-concept coherent caloritronic devices such as heat transistors and thermal splitters could be envisioned which exploit phase-dependent heat transfer peculiar to the Josephson effect.

We ackowledge F. Taddei for fruitful discussions and for a  careful reading of the manuscript.
We thank C. Altimiras, C. W. J. Beenakker, M. Di Ventra,  T. T. Heikkil$\ddot{\text{a}}$, M. A. Laakso, F. Portier, and P. Spathis for comments, 
and the EC FP7 program No. 228464 ``Microkelvin'' for partial financial support.

\section{Methods Summary}

The structures  have been fabricated with e-beam lithography and three-angle shadow-mask evaporation of metals onto an oxidized Si wafer through a suspended resist mask. In the e-beam evaporator, the chip is initially tilted at an angle of $28^{\circ}$, and 20 nm of Al are deposited to realize S$_2$ and S$_3$. The sample is then exposed to 380 mTorr  of O$_2$ for 4.5 minutes to form the SQUID tunnel barriers after which it is tilted to $0^{\circ}$ for the deposition of 25 nm of Al forming S$_1$ as well as heaters and thermometers probes. The chip is subsequently exposed to 800 mTorr of O$_2$ for 4.5 minutes to form heaters, thermometers, source and drain tunnel junctions. Finally, 30 nm of Cu are deposited at $42^{\circ}$ to realize source and drain. 

The magneto-electric characterization of the samples was performed down to 235 mK in a filtered $^3$He cryostat. Current biasing of thermometers was obtained through battery-powered floating sources, whereas voltage and current were measured with room-temperature preamplifiers. Flux-to-temperature transfer functions were measured with low-frequency lock-in technique by superimposing a small modulation to the applied magnetic field.

In the energy-balance equations [see Eqs. (\ref{balance})], $\dot{Q}_{probe}=\dot{Q}_{qp}^{probe}-\dot{Q}_{int}^{probe}$ \cite{Maki,Guttman2,Giazotto,Guttman1, Sauls1,Sauls2},
$\dot{Q}_{qp}^{probe}=\frac{2}{e^2 R_{p}}\int^{\infty}_{0} d\varepsilon \varepsilon \mathcal{N}_1 (\varepsilon)\mathcal{N}_3 (\varepsilon)[f_1(\varepsilon)-f_3(\varepsilon)]$,
$\dot{Q}_{int}^{probe}=\frac{2}{e^2 R_{p}}\int^{\infty}_{0} d\varepsilon \varepsilon \mathcal{M}_1 (\varepsilon)\mathcal{M}_3 (\varepsilon)[f_1(\varepsilon)-f_3(\varepsilon)]$, $\mathcal{N}_3 (\varepsilon)=\mathcal{N}_2 (\varepsilon)$, $\mathcal{M}_3 (\varepsilon)=\mathcal{M}_2 (\varepsilon)$, and $f_3(\varepsilon)=f_2(\varepsilon)$.  
Furthermore, $\dot{Q}_{source}=\frac{2}{e^2 R_{s}}\int^{\infty}_{0} d\varepsilon \varepsilon \mathcal{N}_1 (\varepsilon)[f_{s}(\varepsilon)-f_1(\varepsilon)]$, $\dot{Q}_{drain}=\frac{2}{e^2 R_{d}}\int^{\infty}_{0} d\varepsilon \varepsilon \mathcal{N}_1 (\varepsilon)[f_1(\varepsilon)-f_{d}(\varepsilon)]$ \cite{RMP}, $f_{s(d)}(\varepsilon)=[1+\text{exp}(\varepsilon/k_BT_{source(drain)})]^{-1}$, $\dot{Q}_{ep}=\Sigma \mathcal{V}(T_{drain}^5-T_{bath}^5)$ \cite{RMP,Roukes}, $\Sigma\simeq 3\times 10^9$ WK$^{-5}$m$^{-3}$ is the electron-phonon coupling constant for Cu \cite{RMP}, and $\mathcal{V}\simeq 2\times 10^{-20}$ m$^{3}$ is drain volume.
For the numerical solution of Eqs. (\ref{balance}) we set the structure parameters as extracted from the experiment, and
varied $R_{p}$ between  $\sim 100\%$ and $\sim 125\%$ to match measured data.

\end{document}